\documentclass[
reprint,
superscriptaddress,
 amsmath,amssymb,
 aps,
]{revtex4-1}

\usepackage{comment}
\usepackage{verbatim}
\usepackage{graphicx}
\usepackage{dcolumn}
\usepackage{bm}
\usepackage{hyperref}
\usepackage{inputenc}

\begin{document}

\preprint{APS/123-QED}

\title{Model-free hidden geometry of complex networks}

\author{Yi-Jiao Zhang}
\affiliation{Institute of Computational Physics and Complex Systems, Lanzhou University, Lanzhou, Gansu 730000, China}

\author{Kai-Cheng Yang}
\affiliation{Center for Complex Networks and Systems Research, Luddy School of Informatics, Computing, and Engineering, Indiana University, Bloomington, Indiana 47408, USA}

\author{Filippo Radicchi}
\affiliation{Center for Complex Networks and Systems Research, Luddy School of Informatics, Computing, and Engineering, Indiana University, Bloomington, Indiana 47408, USA}
\email{filiradi@indiana.edu}


\begin{abstract}
The fundamental idea of embedding a network in a metric space is rooted in the principle of proximity preservation. Nodes are mapped into points of the space with pairwise distance that reflects their proximity in the network.  
Popular methods employed in network embedding either rely on implicit approximations of the principle of proximity preservation or implement it by enforcing the geometry of the embedding space, thus hindering geometric properties that networks may spontaneously exhibit. 
Here, we take advantage of a model-free embedding method explicitly devised for preserving pairwise proximity, and characterize the geometry emerging from the mapping of several networks, both real and synthetic.
We show that the learned embedding has simple and intuitive interpretations: 
the distance of a node from the geometric center is representative for its closeness centrality, and the relative positions of nodes reflect the community structure of the network.
Proximity can be preserved in relatively low-dimensional embedding spaces, and the hidden geometry displays optimal performance in guiding greedy navigation regardless of the specific network topology. 
We finally show that the mapping provides a natural description of contagion processes on networks, with complex spatiotemporal patterns represented by waves propagating from the geometric center to the periphery.
The findings deepen our understanding of the model-free hidden geometry of complex networks.
\end{abstract}

\maketitle

\section{Introduction}

A wealth of recent papers from the physics community have demonstrated that representing complex networks in metric space may be beneficial in several respects~\cite{boguna2020network}.
Geometric representations provide intuitive explanations for several properties of real-world networks, including structural features~\cite{Papadopoulos2012Popularity, wu2015emergent, bianconi2017emergent}, navigability~\cite{boguna2009navigability,gulyas2015navigable}, robustness~\cite{kleineberg2017geometric, osat2020k}, community organization~\cite{zuev2015emergence, faqeeh2018characterizing}, and functional modularity~\cite{gallos2012small,de2017diffusion}. 
Also, the computer science community displays a growing interest in embedding networks in vector space~\cite{hamilton2017representation, goyal2018graph, chami2020machine}.
Resulting representations allow for direct applications of standard machine learning algorithms in traditional graph analysis tasks, such as link prediction~\cite{grover2016node2vec}, node classification~\cite{perozzi2014deepwalk}, community detection~\cite{yang2016modularity}, and graph visualization~\cite{Tang2015line}.

The two strains of work share the same basic rationale: nodes in the space should have pairwise distance that reflects their similarity or proximity in the graph. 
However, proximity preservation is implemented in different ways depending on the scientific community of reference.

The approach adopted by the physics community often involves explicit network models.
Such an approach has the advantage of providing an immediate interpretation of the embedding.
One of the most influential models, the so-called
popularity-similarity-optimization
model (PSOM)~\cite{Papadopoulos2012Popularity,papadopoulos2015network_1}, is a very good example of this statement. PSOM assumes that nodes are represented by points in the hyperbolic disk, and that their radial and angular coordinates are representative respectively of individual popularity and pairwise similarity. The model further assumes connection probability between pairs of nodes to be an explicit function of their distance in the hyperbolic space. The embedding of a given network in the hyperbolic disk is then found by fitting the network against PSOM, with the hyperbolic coordinates of the nodes playing the role of the fitting parameters. Model-based approaches to network embedding are intuitive. However, they are useful only as long as the assumed generative model is sufficiently accurate in describing the structure of the fitted network~\cite{boguna2020network}. 

By contrast, typical methods for network embedding developed by the computer science community do not assume explicit generative models~\cite{goyal2018graph, chami2020machine}. 
Such a model-free approach generally relies on defining a metric of pairwise node similarity and then seeking the vectorized representation that best preserves the overall similarity of an observed graph~\cite{qiu2018network,liu2019general}.
The approach is flexible enough to provide a nontrivial geometric representation of any network. 
However, the interpretation of the inferred geometry may not be immediate. Take for example classical methods based on the spectral decomposition of graph operators, such as Laplacian Eigenmaps~\cite{belkin2002laplacian}. 
Only some of the principal components of the space, where the graph is projected to, have an intuitive physical meaning. 
Further the distance between points in the embedding space mostly reflects the similarity of the nodes in terms of common connections to other nodes but is not necessarily related to their physical distance in the graph space. 
The meaning of the geometric representation of a network becomes even more opaque for sophisticated graph embedding methods aiming at preserving \textit{ad hoc} similarity metrics through quite involved optimization techniques, such as Deepwalk~\cite{perozzi2014deepwalk}, node2vec~\cite{grover2016node2vec}, and HOPE~\cite{ou2016asymmetric}. The lack of a simple geometric interpretability of network embeddings doesn't hinder their usefulness in machine learning tasks.
However, it may seriously impede human understanding in applications of graph embeddings on critical issues such as identifying high-risk patients~\cite{choi2017gram} and repurposing drugs for the treatment of novel diseases~\cite{yamanishi2008prediction,gysi2020network}.

\begin{figure*}[!htb]
\begin{center}
\includegraphics[width=0.9\textwidth]{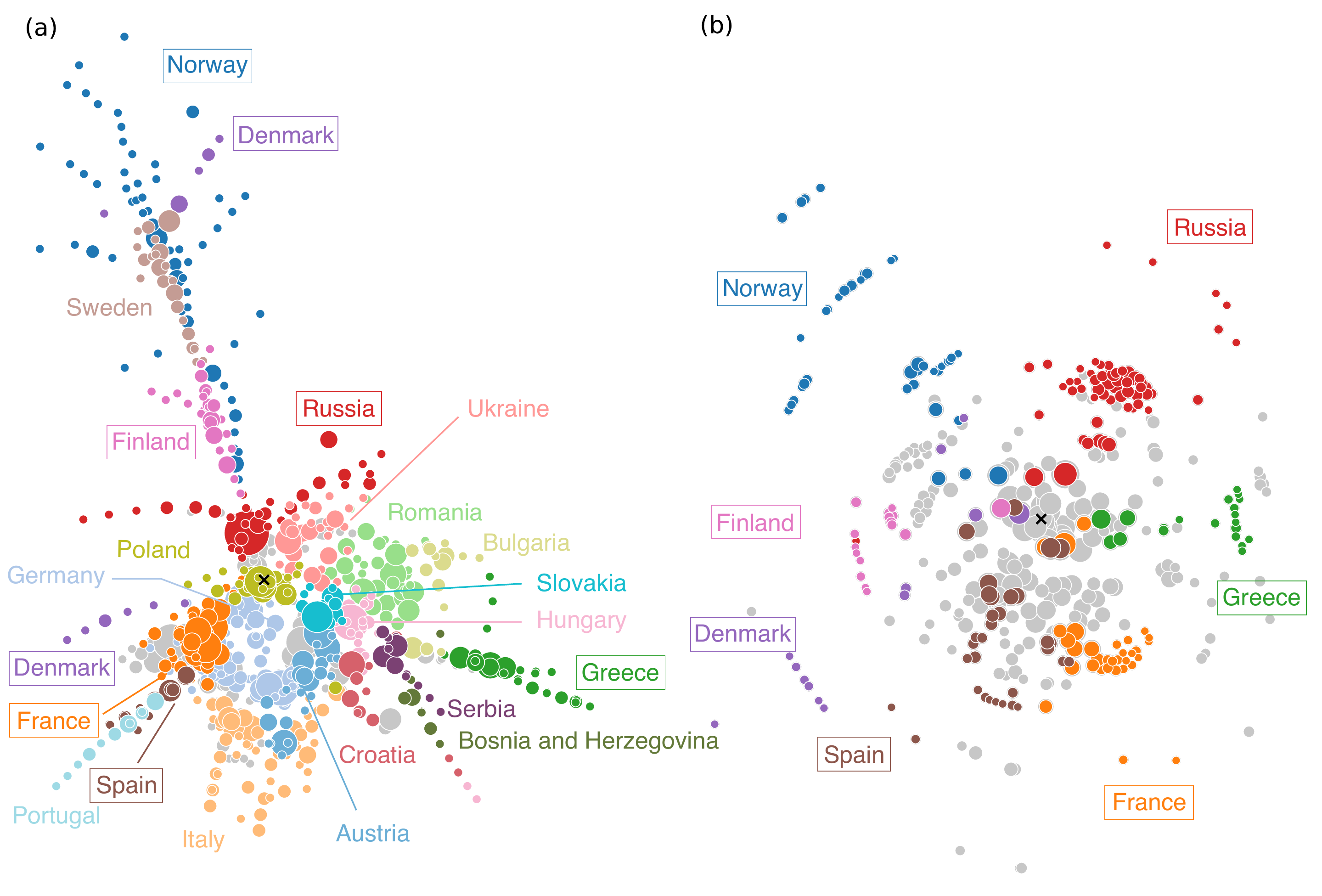}
\end{center}
\caption{\label{fig:europe_2d_map} 
{\bf Model-free maps of real-world networks.} Two-dimensional MDS maps of the (a) European Roads (ER)~\cite{Subelj2011} and (b) European Air Transportation (EAT)~\cite{Guimera2005} networks. For details on the two networks see Appendix~\ref{appendix:networks_2d_map}. The size of nodes in the ER network is proportional to the square of their degrees. For the EAT network, the size of nodes is proportional to the square root of their degrees. Nodes in both figures are colored depending on the country they belong to. For sake of clarity, we label and color only nodes corresponding to a subset of European countries, and display in gray nodes not belonging to this subset.  Also, only nodes with degree centrality larger than one are shown. The black $\textbf{x}$ symbol in both panels denotes the geometric center of the embedding space.}
\end{figure*}

Is it possible to find clear interpretations of the hidden geometry that are learned by model-free embedding approaches?
In this paper, we tackle directly this question. We consider geometric representations where the shortest path length among all pairs of nodes in the original network is preserved as much as possible in the embedding space.
To avoid introducing prior knowledge on the hidden geometry of the network, we use Euclidean space for the embedding.
Other than the objective to find the network representation that best preserves pairwise shortest path lengths, the model-free technique imposes no further restrictions.
We stress that the embedding method considered here already has been employed in Ref.~\cite{pich2009applications} to assist graph drawing, in Refs.~\cite{kuchaiev2009geometric, you2010using, cannistraci2013minimum} for the prediction of missing protein-protein interactions in biological networks, and in Ref.~\cite{muscoloni2017machine} to aid the embedding of networks in the hyperbolic space. Here, we reuse the technique because it represents the most natural choice if the goal is providing embeddings that are congruent with the network structure with no underlying models and minimum restrictions.
The emerging geometries are network dependent, and not necessarily similar to previously suggested ones~\cite{Krioukov2010Hyperbolic}. Learned geometries have simple and intuitive meanings. For example, the distance of a node from the geometric center is representative of its closeness centrality~\cite{lu2016vital}, while the relative positions of nodes reflect network community structure~\cite{fortunato2010community}.
We find that graph distance can be very well preserved in relatively low-dimensional embedding spaces.
Our analysis further shows that the model-free embedding can assist network routing~\cite{kleinberg00,boguna2009navigability,gulyas2015navigable} with high performance across a wide spectrum of real and synthetic networks, including the ones that the model-based approach fails on.
The structure of the model-free embedding also suggests a geometrical representation of spreading processes, which corresponds to waves propagating from the center to the periphery of the embedding space.

\section{results}
\subsection{Inferring the model-free embedding map}

The approach to network embedding that we consider here is a direct implementation of the principle of proximity preservation: graph distance, i.e., the length of the shortest path connecting two nodes in the network, should to be preserved in terms of metric distance, i.e., the length of the shortest path connecting the projections of the two nodes in the embedding space. To this end, we take advantage of MultiDimensional Scaling (MDS, see Appendix~\ref{appendix:mds} for details), a dimensionality reduction method specifically devised to translate information about the pairwise distances among a set of $N$ objects into a configuration of $N$ points mapped into an abstract Euclidean space~\cite{mead1992review, borg2005modern}.
For networks with more than 7,000 nodes, we deploy the landmark MDS~\cite{Silva03globalversus,silva2004sparse} to approximate the model-free embedding (see Appendix~\ref{appendix:mds} for details).
As do many other machine learning techniques, MDS also has a relatively long history and is now experiencing renovated interest in the emerging field of data science~\cite{young2013multidimensional}. MDS has been considered in a broad set of disciplines, such as psychology, sociology, economics, biology, chemistry, and archaeology~\cite{borg2005modern}.

As we mentioned above, some applications of MDS to network data exist~\cite{pich2009applications, kuchaiev2009geometric, you2010using, cannistraci2013minimum, muscoloni2017machine, bertagnolli2019network}. Also, we remark that the embedding method is quite often referred to as ISOmap rather than MDS, e.g., in Refs.~\cite{you2010using, cannistraci2013minimum, muscoloni2017machine,chami2020machine}. 
ISOmap is a method for dimensionality reduction of data points in arbitrary space that relies on MDS~\cite{tenenbaum2000global}. It consists of three steps: (1) generation of a network representation of the data, (2) computation of the shortest path length matrix of the obtained network representation, (3) application of MDS embedding. In our case, the network is already at our disposal, so no data preprocessing is needed. We prefer to think that what we use is the MDS method with input consisting of the shortest path length matrix of a network. Therefore, we will refer to the embedding method used in this paper as MDS.


\subsection{Illustration and interpretation of the model-free embedding map}

To illustrate the model-free hidden geometry of networks learned by MDS in an intuitive setting, in Fig.~\ref{fig:europe_2d_map} we display two-dimensional maps for two real-world networks: the European Roads (ER)~\cite{Subelj2011} and European Air Transportation (EAT)~\cite{Guimera2005} networks. Nodes in both networks represent main European cities, with the caveat of large cities potentially represented by multiple airports in the EAT network (see details in Appendix~\ref{appendix:networks_2d_map}).
In general, nodes belonging to the same European country are mapped close to each other in both maps, so that there is a positive correlation between distance in the embedding space and physical distance between pairs of cities. Some exceptions to this rule are apparent. For example in the ER map, Denmark nodes are separated into two groups corresponding to the Jutland and Zealand regions of the country. These groups are respectively mapped close to Germany or Sweden, reflecting actual network proximity rather than purely geographical one.
Even though all nodes represent European cities, the emerging geometrical pattern of the two maps is rather different because of the different ways that connections are drawn in the networks.
In the ER network, geographical constraints affect the existence of network connections between pairs of cities, and thus geographically adjacent countries are adjacent in the MDS map too. Also, the relative location of countries is congruent with their geographical position. By contrast, the absence of physical constraints in establishing connections among cities in the EAT network leads to a circular geometrical pattern, resembling the one imposed in hyperbolic embedding~\cite{boguna2009navigability}. Cities are staggered into neat shells denoting their network centrality, and clustered into narrow angular slices depending on the country they belong to.

\begin{figure}[!htb]
\begin{center}
\includegraphics[width=0.4\textwidth]{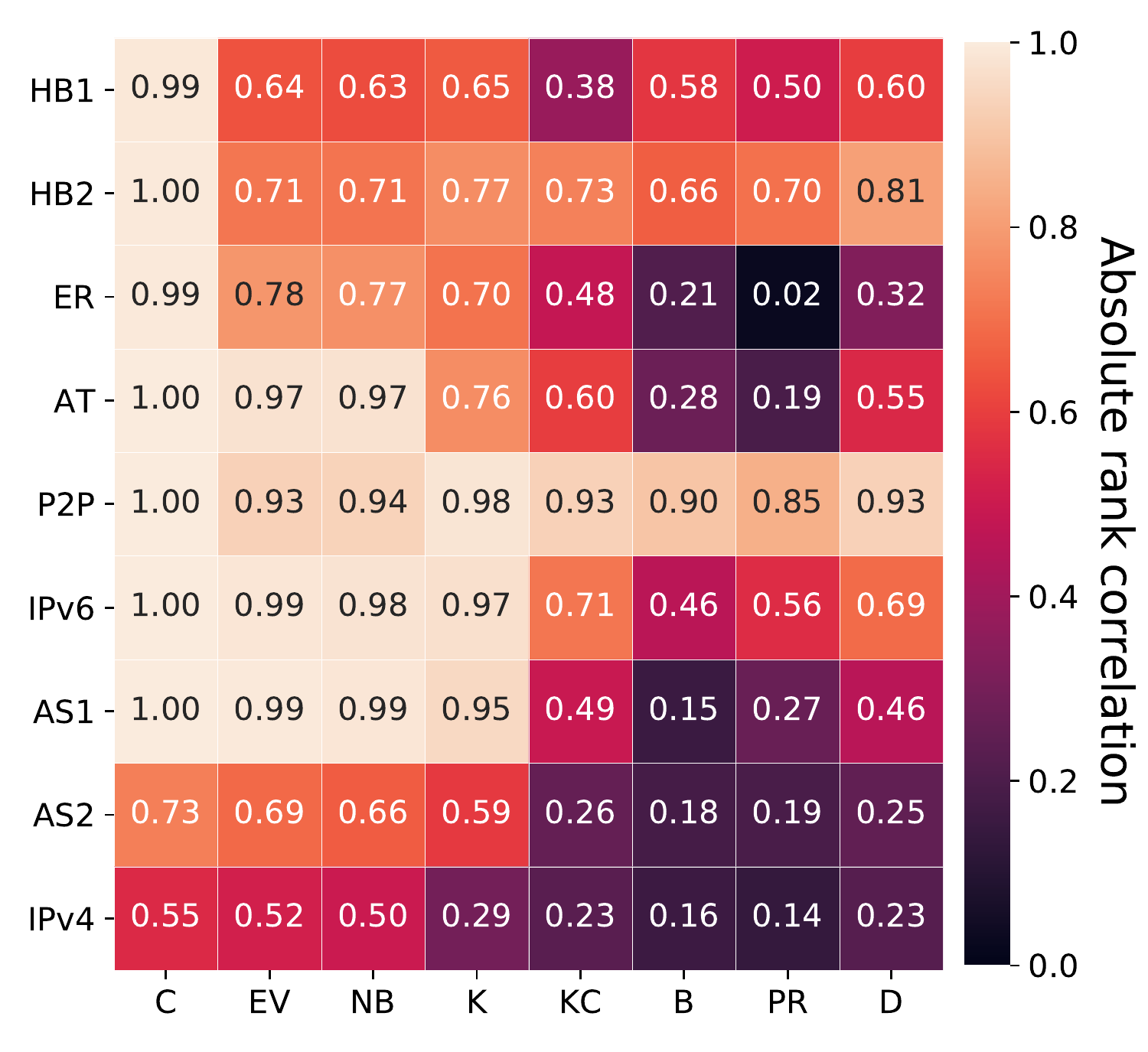}
\end{center}
\caption{\label{fig:compare_centralities} 
{\bf Geometric interpretation of network centrality metrics.}
Pairwise rank correlation between different network centrality metrics and 
the distance from the geometric center of the MDS embedding.
Here, the embedding dimension is $d = 100$. 
Note that we take absolute value of correlation coefficients for convenient comparison. We compute the correlation coefficients for nine real-world networks (see Table~\ref{tab:network_information} for details). 
Network centralities considered  are closeness (C), eigenvector (EV), non-backtracking (NB), Katz (K), k-core (KC), betweenness (B), PageRank (PR), and degree (D).}
\end{figure}

Now that we have an intuitive picture of the resulting network geometry, we can proceed with the interpretation and characterization of it. 
As we already observed while describing Fig.~\ref{fig:europe_2d_map}, a possible geometric pattern emerging from the mapping is characterized by a hyper-spherical organization of the nodes around the geometric center. 
To explore potential geometric interpretations of network properties, we calculate rank correlations between the distance of the nodes from the geometric center in the hidden space and observable network centrality metrics~\cite{lu2016vital}, including degree, betweenness~\cite{freeman1977set},  eigenvector~\cite{bonacich1972factoring}, Katz~\cite{katz1953new}, PageRank~\cite{brin1998anatomy}, nonbacktracking~\cite{martin2014localization}, k-core~\cite{kitsak2010identification}, and closeness~\cite{sabidussi1966centrality} centralities (see results in Fig.~\ref{fig:compare_centralities} and Appendix.~\ref{appendix:relation_to_centrality}). 
We find that the distance from the center of the embedding always shows a highly negative correlation with the closeness centrality.
Further analysis suggests the existence of a natural inequality between network closeness centrality and distance from the MDS geometric center in the case of perfect embedding (see Appendix.~\ref{appendix:relation_to_centrality}).
The model-free geometry provides an elegant and intuitive interpretation for the closeness centrality in the hidden space.
The finding also emphasizes a major difference of the model-free embedding with hyperbolic embedding, where the radial coordinate of a node is by definition a decreasing function of its degree centrality.

\subsection{Ability to preserve network information}

Previous studies dealing with the MDS mapping of networks focus on embedding spaces with two or three dimensions~\cite{pich2009applications,kuchaiev2009geometric, you2010using,muscoloni2017machine}. In the study of geometrical properties of networks, we do not have to deal with a fixed value of the dimension for the embedding space. Rather, we can use it as a free parameter that allows us to find the right compromise between the level of reduction of network information and the effectiveness of the embedding in preserving graph distance.
What is the minimum number of dimensions required to achieve a reasonable level of congruence between graph and embedding distance? The answer to this question should be network dependent. For a network with strong geographical constraints such as the ER network, a two-dimensional space should suffice to provide an accurate geometrical description of the network. For other networks, however, we may need a higher number of dimensions. 

We note that the authors of Ref.~\cite{cannistraci2013minimum} studied how performance of link prediction in protein-protein interaction networks is affected by the choice of the embedding dimension of the MDS space. Here, we differentiate from such an analysis in several important respects. First, we focus on the quality of the embedding itself and not a specific downstream task, e.g., link prediction. This fact gives greater generality to our analysis. Second, we systematically study the quality of MDS embedding of synthetic networks, so that we are allowed to vary system size while keeping other properties invariant. The analysis serves to understand how the dimension of the underlying space should be chosen for a given network size. Third and most important, we consider different types of real-world networks, not just biological ones. In particular, the quality of the embedding at a certain dimension depends on the size of a real-world network in a similar way as observed for synthetic networks. 

\begin{figure}[!htb]
\begin{center}
\includegraphics[width=0.48\textwidth]{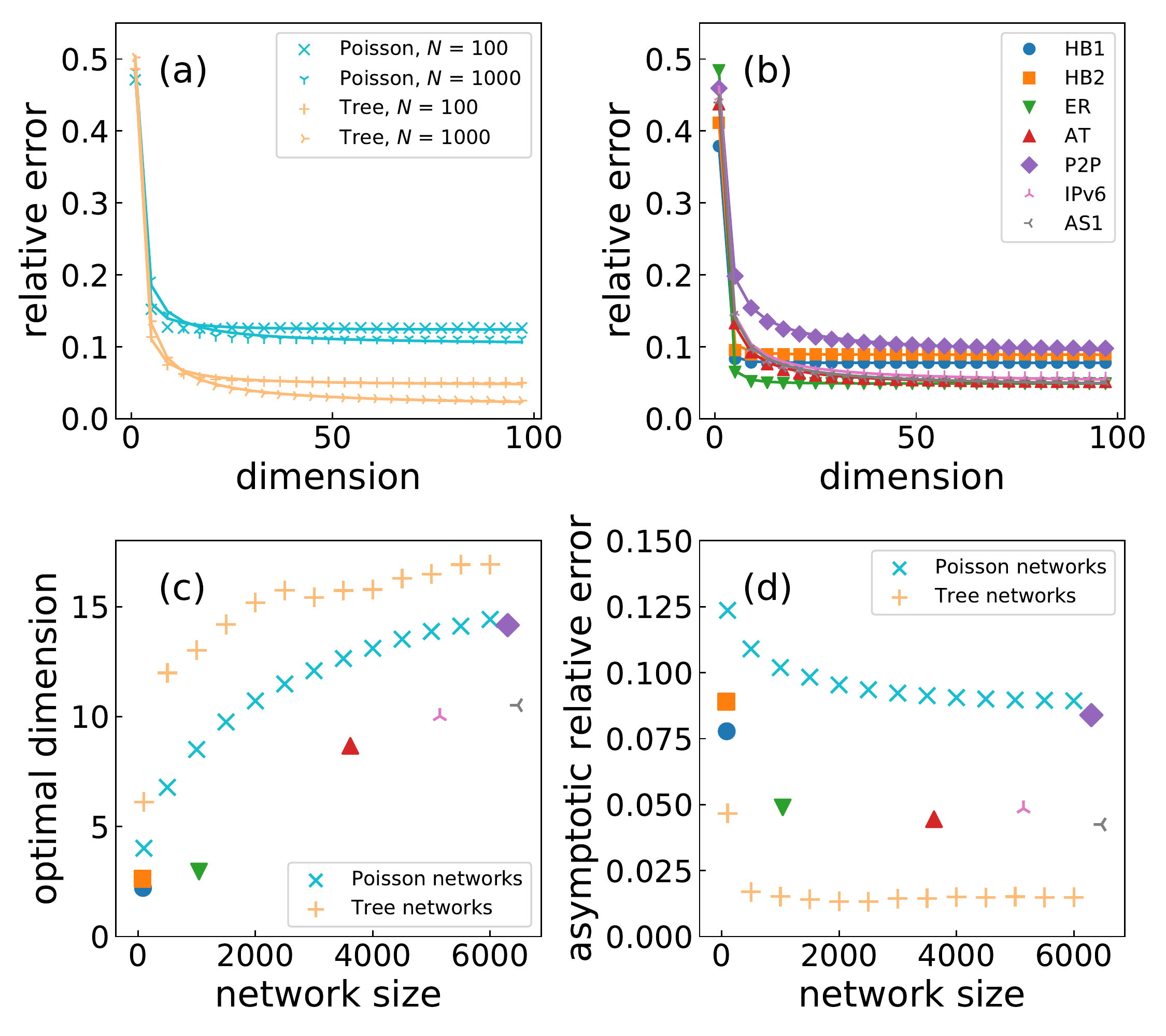}
\end{center}
\caption{\label{fig:performance_mds} 
{\bf Preservation of graph distance in the embedding space.} (a) Relative error between the shortest path length in the graph and the distance in the embedding space as a function of the embedding dimension. Relative error is averaged over all pairs of nodes in the network (see Appendix~\ref{appendix:relative_error} for details). We consider different types of network models, either Poisson networks with average degree $\langle k \rangle =4$ or 
$m$-ary trees with $m = 3$, and two different network sizes $N = 100$ and $N = 1,000$. Continuous curves in the plot are obtained by fitting data points with power-law decaying functions towards an asymptotic relative error value (see Appendix~\ref{appendix:optimal_dimension}). (b) Same as in panel (a), but for real-world networks (see Table~\ref{tab:network_information}). (c) Optimal embedding dimension as a function of the network size $N$ (see Appendix~\ref{appendix:optimal_dimension}). For network models, we consider different sizes. Real-world networks are denoted by a single point in the plot. (d) Asymptotic value of the relative error as a function of the network size.}
\end{figure}

\begin{table*}[!htb]
\caption{
{\bf Model-free embedding of real and synthetic networks.} We summarize here
basic information and selected metrics of the model-free embedding of the real-world networks and synthetic networks considered in this paper. 
From left to right, we report name and acronym of the network, size $N$ of the network, total number of edges $L$, average degree $\textrm{$\langle k \rangle$}$, asymptotic relative error $\hat{E}_{\infty}$, optimal embedding dimension $\hat{d}_{o}$ obtained at accuracy level $\epsilon = 0.05$, the success rate $p_s$ of greedy routing and the rank correlation coefficient $\rho$ between the distance from the geometric center in the MDS space and the average infection arrival time of nodes.
The standard deviations of $\hat{E}_{\infty}$ and $p_s$ in multiple repetitions are smaller than $0.002$ and $0.007$, respectively, for all networks.
The $p$ values of $\rho$ are all close to zero.
For IPv4 and AS2, the values of $\rho$ are not available due to computational limitations and other metrics are obtained using landmark MDS (see Appendix~\ref{appendix:mds}).
Three instances of PSOM use the same parameters with $N=5,000$, $\gamma = 2.1$, and $\langle k \rangle=5$ but different values of $T$.
Only the largest connected component is kept for each instance.
Since the PSOM instances are introduced to test greedy routing specifically, the values of $\rho$ are not provided.
}
\label{tab:network_information}
\begin{ruledtabular}
\begin{tabular}{rrrccccc}
 & $N$ & $L$ & \textrm{$\langle k \rangle$} & $\hat{E}_{\infty}$ & $\hat{d}_{o} (\epsilon = 0.05) $ & $p_s$ & $\rho$\\
\hline
\textbf{Real-world networks} &&&&&&& \\
\hline
Human brain, layer1 (HB1) \cite{Kleineberg2016Hidden} & 85 & 230 & 5.412 & 0.078 & 2 & 1.000 & 0.818\\
\hline
Human brain, layer2 (HB2) \cite{Kleineberg2016Hidden} & 78 & 218 & 5.590 & 0.089 & 3 & 1.000 & 0.900\\
\hline
European Road (ER) \cite{Subelj2011} & 1,039 & 1,305 & 2.512 & 0.049 & 3 & 0.924 & 0.967\\
\hline
Air Transportation (AT) \cite{Guimera2005} & 3,618 & 14,142 & 7.818 & 0.045 & 9 & 0.904 & 0.859\\
\hline
P2P (P2P) \cite{ripeanu2002mapping} & 6,299 & 20,776 & 6.597 & 0.084 & 14 & 0.973 & 0.985\\
\hline
AS Oregon Internet (AS1) \cite{Leskovec2005Graphs}& 6,474 & 12,572 & 3.884 & 0.042 & 11 & 0.903 & 0.785\\
\hline
IPv6 Internet (IPv6) \cite{Kleineberg2016Hidden} & 5,143 & 13,446 & 5.229 & 0.049 & 10 & 0.943 & 0.848\\
\hline
IPv4 Internet (IPv4) \cite{Kleineberg2016Hidden} & 37,542 & 95,019 & 5.062 & 0.100 & 38 & 0.912 & -\\
\hline
AS Internet (AS2) \cite{boguna2010sustaining} & 23,748 & 58,414 & 4.919 & 0.077 & 47 & 0.933 & - \\
\hline
\textbf{Synthetic networks} &&&&&&& \\
\hline
Poisson network, $\langle k \rangle=4$ & 1,000 & 2,029 & 4.058 & 0.102 & 9 & 1.000 & 0.986 \\
\hline
$m$-ary tree, $m=3$ & 1,000 & 999 & 1.998 & 0.015 & 13 & 0.913 & 0.879 \\ 
\hline
PSOM, $T = 0.1$ (PSOM1) \cite{Papadopoulos2012Popularity,papadopoulos2015network_1} & 4,521 & 12,175 & 5.386 & 0.027 & 12 & 0.905 & - \\
\hline
PSOM, $T = 0.5$ (PSOM2) \cite{Papadopoulos2012Popularity,papadopoulos2015network_1}& 4,731 & 12,553 & 5.307 & 0.042 & 9 & 0.861 & - \\
\hline
PSOM, $T = 0.9$ (PSOM3) \cite{Papadopoulos2012Popularity,papadopoulos2015network_1}& 4,068 & 6,821 & 3.353 & 0.048 & 9 & 0.877 & - \\
\end{tabular}
\end{ruledtabular}
\end{table*}

We measure the quality of a network representation by estimating the average value of the relative error between the distances in the graph and the embedding spaces across all pairs of nodes in the network (see details in Appendix~\ref{appendix:relative_error}).
As the results of Fig.~\ref{fig:performance_mds} show, the relative error committed by MDS embedding in preserving graph distance quickly saturates to an asymptotic value as the dimension of the embedding space increases. The phenomenon is apparent for all networks we consider, either real and synthetic. Ideally, the asymptotic value of the relative error should equal zero for a perfect embedding. We observe, however, that perfect MDS embedding cannot be achieved even in high-dimensional spaces. As a matter of fact, the shortest path length between nodes exhibits persistent triangle inequality violation when we try to embed the network into Euclidean space, thus imposing a limit to the quality of the resulting embedding~\cite{chen2011phoenix, bourgain1985lipschitz, linial1995geometry}. 
We stress that the asymptotic relative error values, although strictly larger than zero, are rather small (see Table~\ref{tab:network_information}). Considering the fact that the real-world networks we consider are very heterogeneous in terms of size, edge density, degree distribution, etc., the quality of the MDS representations that we obtain is excellent for sufficiently high-dimensional embedding spaces. To quantitatively establish an optimal value for the embedding dimension that balances efficiency and quality of the representation, we adapt the parametric method of Ref.~\cite{gu2020defining} to our error metric (see Appendix~\ref{appendix:optimal_dimension}). Our empirical results show that the optimal dimension grows only logarithmically with the network size in synthetic network models. For the real networks considered in this paper, optimal dimension values display a similar behavior as in synthetic networks. In particular, all networks have optimal embedding dimension smaller than $50$ (see Table~\ref{tab:network_information}). 

We also evaluate the ability of MDS to preserve graph distance with the Pearson correlation coefficients. The results are qualitatively similar with relative error (see Appendix~\ref{appendix:pearson_correlation}).

\subsection{Efficiency in guiding routing}

To further validate the quality of the model-free geometry learned from MDS embedding, here we test its ability of guiding navigation on networks through greedy routing.
This is a quite important task in several contexts, including routing of information packets on the Internet, diffusion of electric signals in neural networks, and the flow of people or goods in transportation and delivery networks~\cite{boguna2009navigability,boguna2010sustaining}. We stress that greedy routing is very different from the downstream tasks usually considered in the validation of other model-free network embedding methods~\cite{hamilton2017representation, goyal2018graph, chami2020machine}. It is rather a common test bed for model-based methods, such as hyperbolic embedding~\cite{boguna2009navigability, boguna2010sustaining, muscoloni2017machine}. 
 
In greedy routing~\cite{kleinberg00}, a packet should be delivered from the source node $s$ to the target node $t$.  The packet can walk along one edge each at a time until it reaches its destination, preferably following the shortest path connecting $s$ to $t$. If one relies on complete information about the network structure, where every single node is aware of its graph distance from any other node, then the protocol is optimal in the sense that all packets are delivered with $100\%$ success probability and always along the actual shortest path. However, such a procedure requires nodes to store routing tables based on complete network information, and it is thus not scalable. 
Network embedding methods, especially the MDS embedding considered here, provides a natural solution, as the distance between nodes in the embedding space is representative of distance in the graph space.
While walking towards its target $t$, a packet sitting on the arbitrary node $i$ may simply move to the neighbor $j$ of node $i$ that is closest to $t$ according to their distance in the embedding space. The protocol requires each individual node to store information only about its neighbors' coordinates in the space, thus making it highly scalable. 

Clearly under the MDS protocol, some packets may visit the same node twice. In that case, the packet is considered lost and the routing process unsuccessful. Also, even if a packet is delivered, it may have not followed the true shortest path connecting the source node to the target. Success rate~\cite{boguna2009navigability,boguna2010sustaining} and efficiency~\cite{faqeeh2018characterizing} for randomly chosen pairs of source and target nodes are the standard metrics of performance for the evaluation of this type of navigation protocols (see Appendix~\ref{appendix:greedy_routing} for details).

\begin{figure}[!htb]
\begin{center}
\includegraphics[width=0.48\textwidth]{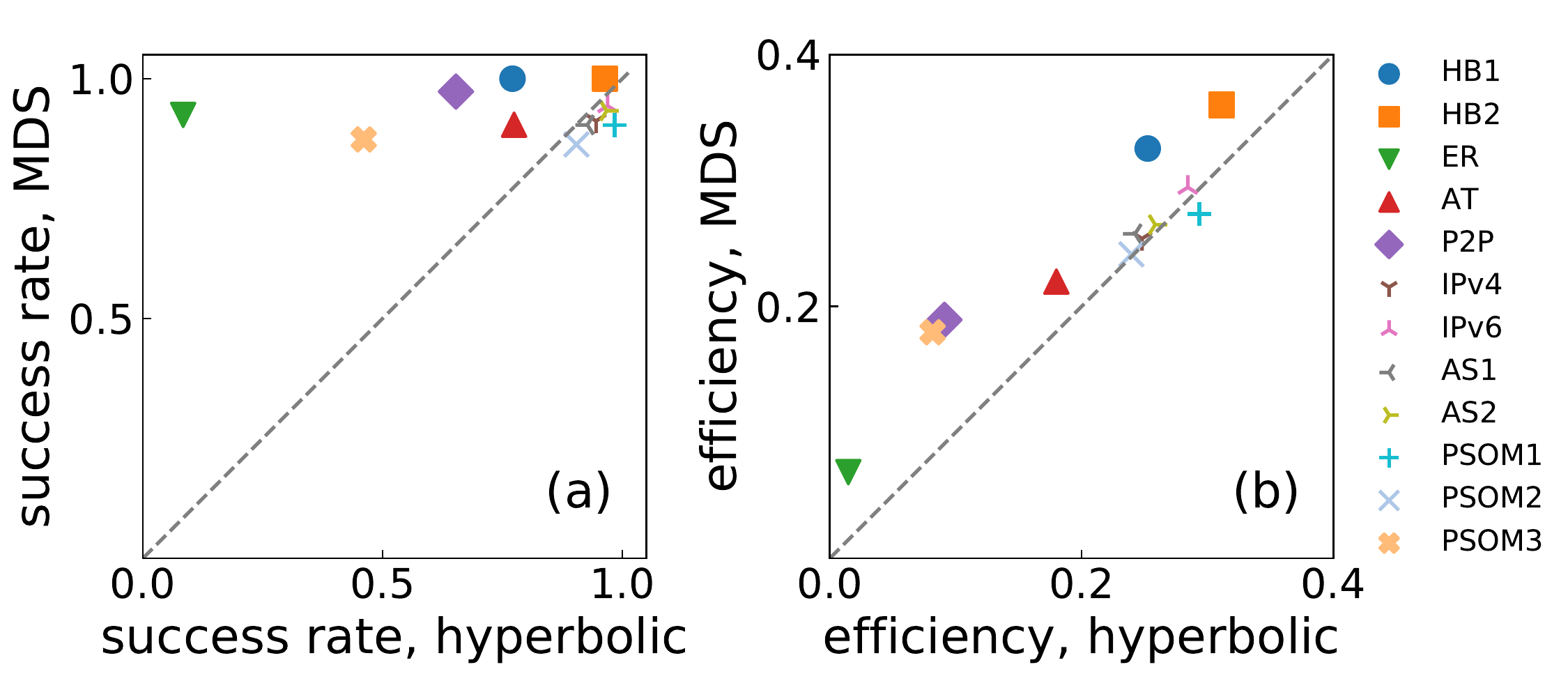}
\end{center}
\caption{\label{fig:comparison_greedy_routing_rate} 
{\bf Routing protocols relying on network geometry.}
We compare (a) success rate and (b) efficiency 
of greedy routing in MDS and hyperbolic spaces for different real-world networks (acronyms and symbols are the same as in Fig.~\ref{fig:performance_mds}). We consider also three instances of PSOM with $N = 5000, \langle k \rangle = 5$ and degree exponent $\gamma = 2.1$, and different values of the temperature parameter $T = 0.1, T = 0.5, T = 0.9$ (respectively indicated as PSOM1, PSOM2, and PSOM3). The markers located in the upper triangle area represent the cases where success rate or efficiency in MDS space is higher than in hyperbolic space. }
\end{figure}

We test MDS greedy routing on nine real-world networks and three instances of the PSOM (see Table~\ref{tab:network_information} and Appendix~\ref{appendix:networks} for details).
We use greedy routing based on hyperbolic embedding as the term of comparison to assess the performance of the MDS protocol, see Fig.~\ref{fig:comparison_greedy_routing_rate}. The hyperbolic embedding algorithm we use here is HyperMap-CN~\cite{papadopoulos2015network_1}, one of the best among similar methods considered in Ref.~\cite{muscoloni2017machine}.
Regardless of the network, the success rate of the MDS greedy protocol is excellent. Hyperbolic embedding generates protocols with performance comparable to MDS only in some networks, likely those networks that are suitably described by hyperbolic geometry, while it provides much poorer performance than MDS in many other networks.
The GR-score~\cite{muscoloni2017machine, muscoloni2019navigability} is also calculated to measure the performance of greedy routing. The results are the same as for the success rate (see Appendix~\ref{appendix:greedy_routing}).

\subsection{Application in mapping the contagion spreading on networks}

In this final section, we provide evidence that MDS embedding may be used for the geometric description of the spatiotemporal patterns of spreading process occurring on networks~\cite{pastor2015epidemic}. 
Specifically, we hypothesize that MDS may provide an effective representation of complex spreading patterns in terms of standard wave equations in continuous Euclidean space. Such a representation would have the great potential of allowing us to leverage knowledge of the mathematics of waves in classic systems and eventually exploring the analog of wave-related physical phenomena, such as interference and resonance, in spreading processes occurring on networks. 
The rationale of such an hypothesis stems from the fact that infection arrival time can be approximated by network centrality measures~\cite{borgatti2005centrality,moore2020predicting}, and that the distance to the embedding center of each node corresponds to its closeness centrality.


\begin{figure}[!htb]
\begin{center}
\includegraphics[width=0.48\textwidth]{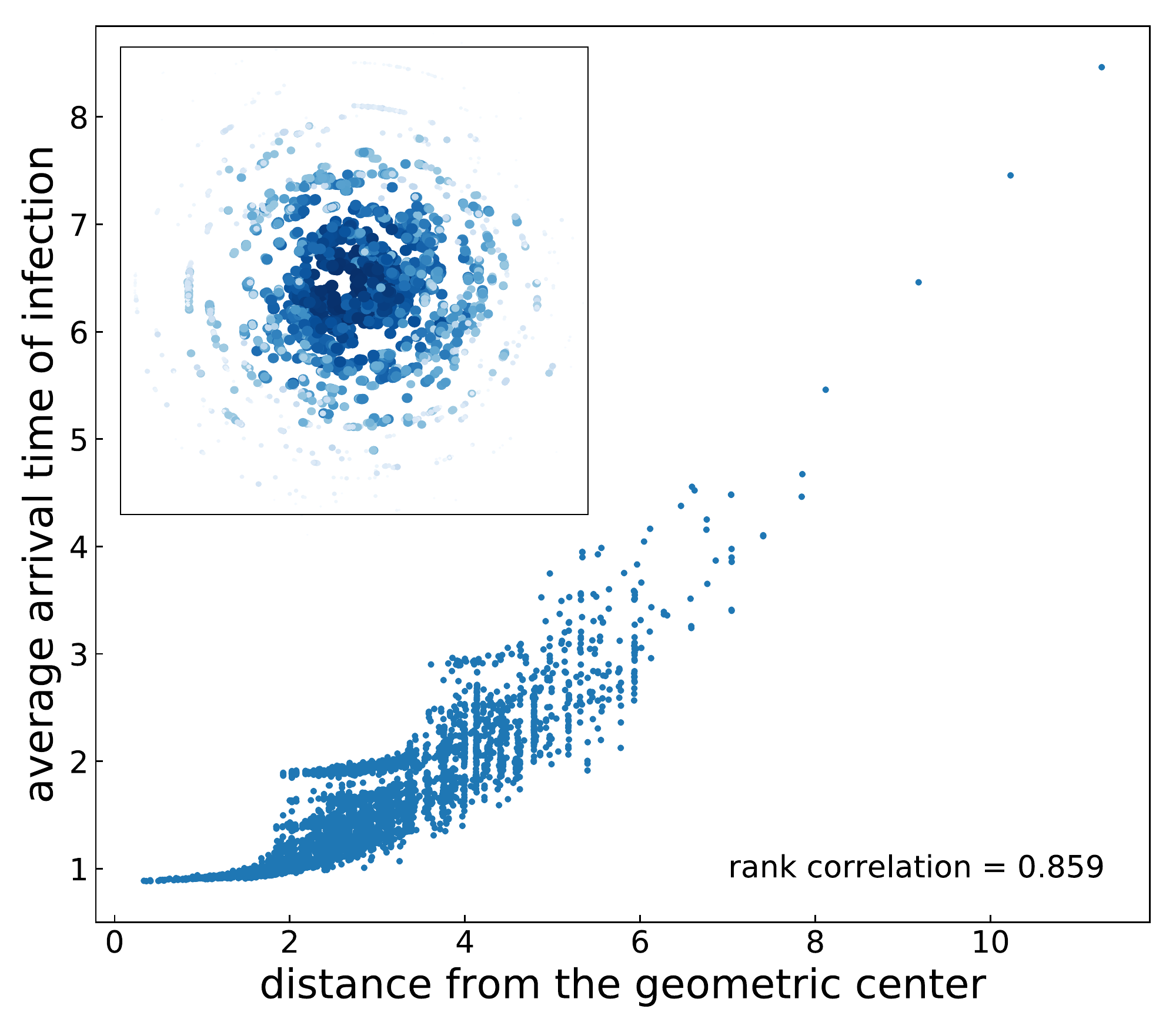}
\end{center}
\caption{\label{fig:time_distance_relation_AT} 
{\bf Geometric description of spreading in complex networks.} In the main panel, we display the scatter plot of the average infection arrival time against the distance of nodes from the geometric center of the MDS map. Each point in the plot is a node of the network. The average infection arrival time is calculated based on $N$ independent instances of the susceptible-infected model simulated on the air transportation (AT) network with a random seed each time, where $N$ is network size. The network is embedded in  $d=100$ dimensions.  For illustrative purposes only, we display in the inset what spreading looks like in a two-dimensional MDS map. Nodes with larger size and darker color are infected earlier.
}
\end{figure}

To test our hypothesis, we consider the susceptible-infected (SI) model as the spreading process (see Appendix~\ref{appendix:si_model} for details). We simulate it on different real-world and synthetic networks. In all our simulations, we always start from a configuration where all nodes are in the susceptible state, with the exception of one node in the infected state. We measure the time required for the infection to arrive to each node. The source of the infection is randomly chosen, and results are averaged over a number of instances equal to the size of the network $N$.
Figure~\ref{fig:time_distance_relation_AT} shows the relationship between the average infection arrival time of each node and its distance to the geometric center for the AT network. The high value of the rank correlation denotes that spreading is well represented by a wave whose front is moving in a rather homogeneous way in the embedding space. The finding is further substantiated by the graphical illustration of the inset of Fig.~\ref{fig:time_distance_relation_AT}, where a clear concentric wave appears in the two-dimensional representation of the network. Similar results can be observed for all the other networks (see the correlation values in Table~\ref{tab:network_information} and more results in Appendix.~\ref{appendix:si_process}).
The high correlations between the infection arrival time and the distance from the MDS embedding center across all networks confirm our hypothesis.
We stress that similar or even slightly higher correlation can be measured if closeness centrality is used in place of the distance from the geometric center of the embedding. However, closeness centrality doesn't provide us with an immediate geometric representation of the network, and the description of spreading in terms of wave equations becomes more complicated than the one obtained in the MDS embedding, thus Euclidean, space. Unlike other efforts to reorder the nodes of a network to construct wavelike patterns~\cite{brockmann2013hidden,hens2019spatiotemporal}, the connection between the hidden geometry and contagion spreading here is a natural implication of the model-free hidden geometry of the networks.

\section{Discussion}

Different approaches to the embedding of network in hidden geometric space exist. The model-based approaches from the physics community can offer an immediate interpretation of the learned geometry, but work only when the model and the network topology are congruent. The model-free approaches adopted by the computer science community have no such limitation, but the lack of an immediate interpretation greatly hinders the human understanding of the learned geometry. 
Here, we consider a compromise between the two above approaches, consisting of a model-free method with immediate geometric interpretation. The method, called MultiDimensional Scaling (MDS) relies on the preservation of the shortest path distance in the embedding space. MDS has been considered previously as a viable method to embed networks in applications such as link prediction and graph drawing. Here, we reconsider it as one of the most natural way of learning the hidden geometry of a network, and translate network properties in geometric space. Indeed, we show that the distance of a node from the geometric center corresponds to its closeness centrality, and the relative positions of nodes reflect the network community structure.

Our work shows that the MDS mapping of networks is a meaningful operation, as the distance among pairs of nodes in a network can be preserved to a great extent in relatively low-dimensional vector spaces, 
irrespective of the specific network considered.
Furthermore, the performance of greedy routing on all kinds of networks embedded in the vector space using MDS is better than, or comparable to, the networks mapped using the state-of-the-art hyperbolic embedding method. The MDS embedding space provides an alternative choice of the navigable hidden space for networks, especially for the networks whose structures are not congruent with hyperbolic maps.  

Also, we show that the complex spreading patterns in the original network can be mapped to a propagating wave in the MDS embedding space.
Our results suggest that MDS embedding can be an effective tool to study dynamical process on networks.

In this work, we use Euclidean space as the target and geodesic distance, the most common definition of graph distance, as the input metric to be preserved. Both choices have been made for clarity and simplicity. However,
the approach can be easily generalized to other metric spaces and definitions of pairwise distance among graph nodes. 
For example, it is possible to preserve the geodesic distance in hyperbolic space to take advantage of the hyperbolic geometry, or
it is possible to replace shortest path distance with another metric of 
distance or similarity, e.g., communicability~\cite{estrada2012communicability}, to obtain slightly different MDS embeddings.
One can also unitize weighted path length in the embedding process to incorporate extra information like traffic~\cite{brockmann2013hidden} or characteristics of the dynamics~\cite{hens2019spatiotemporal} for better predictions of the arrival time of infection. 
We leave such extensions for future work.

\acknowledgements

The authors thank C.V. Cannistraci for critical comments on an early version of the manuscript. 
Y.-J.Z. acknowledges support from China Scholarship Council (No.201906180029). 
Y.-J.Z. and F.R. acknowledge partial support from the National Science Foundation (CMMI-1552487). F.R. acknowledges partial support from the U.S. Army Research Office (W911NF-16-1-0104).

\appendix
 
\section{MultiDimensional Scaling (MDS) embedding of networks}
\label{appendix:mds}

\subsection{Algorithm}

We take the geodesic distance matrix $\textbf{D}$ of a network and the embedding dimension $d$ as inputs of the MDS method. The output are the coordinates $\textbf{x}_i = (x_i^{(1)}, \ldots, x_i^{(v)}, \ldots, x_i^{(d)})$ of all nodes $i$ in a $d$-dimensional space. Coordinates are learned from the input data through the minimization of the stress function

\begin{equation}
S(\textbf{x}_1,\textbf{x}_2, ..., \textbf{x}_N|\textbf{D}) = \sum_{i > j} [D_{ij} - ||\textbf{x}_i - \textbf{x}_j||]^2 \; ,  \label{metric_MDS_stress_function}
\end{equation}
where $||\textbf{x}_i - \textbf{x}_j|| = \sqrt{\sum_{v=1}^d (x_i^{(v)} - x_j^{(v)})^2}$ is the Euclidean distance between the points $\textbf{x}_i $ and $\textbf{x}_j$ in the vector space.
The Scaling by Majorizing a Complicated Function (SMACOF)~\cite{borg2005modern} algorithm is used to solve the stress minimization problem. We remark that the choice of the matrix $\mathbf{D}$ to be the geodesic distance between all pairs of nodes is a specific choice made in this paper. One can replace that matrix with any other arbitrary node distance matrix. Also, the minimization of the stress function of Eq.~(\ref{metric_MDS_stress_function}) is the so-called metric MDS, i.e., one of the most popular algorithms within the MDS family. Similarly, the Euclidean distance in Eq.~(\ref{metric_MDS_stress_function}) can be replaced with another distance metric.

The above algorithm does not scale well with network size. 
However, there are many speed-up methods available.
Here we adopt the so-called landmark MDS method~\cite{Silva03globalversus, silva2004sparse}.
The approximation relies on a small set of landmark nodes to embed the other nodes on the basis of distance-based triangulation.
The steps for landmark MDS are the following:
\begin{itemize}
\item[1.] Select $l$ ($l \ll N$) landmark nodes randomly from all nodes of the network. In this paper, we set $l = 1,000$.

\item[2.] Apply metric MDS to find the coordinates of landmarks $\textbf{X}_{\text{lands}}$.

\item[3.] The coordinates of the remaining nonlandmark nodes $\textbf{X}_{\text{non-lands}}$ are computed using the distance-based triangulation, 

\begin{equation}
    \textbf{X}_{\text{non-lands}} = -\frac{1}{2} \textbf{X}_{\text{lands}}^{\dagger} \textbf{B} \; ,
\end{equation}
where $\textbf{X}_{\text{lands}}^{\dagger}$ is the pseudo-inverse of $\textbf{X}_{\text{lands}}$ and the ($i, j$)-entry of $\textbf{B}$ is computed as
\begin{equation}
    B_{ij} = F_{ij}^2 - \frac{1}{l} \sum_{j=1}^l E_{ij}^2 \; ,
\end{equation}
with the matrix $\textbf{E} \in \mathbb{R}^{l \times l}$ representing the distance matrix of the landmark nodes and $\textbf{F} \in \mathbb{R}^{l \times (N - l)}$ the distance between the landmark and nonlandmark nodes.

\end{itemize}

\begin{figure*}[!htb]
\includegraphics[width=0.85\textwidth]{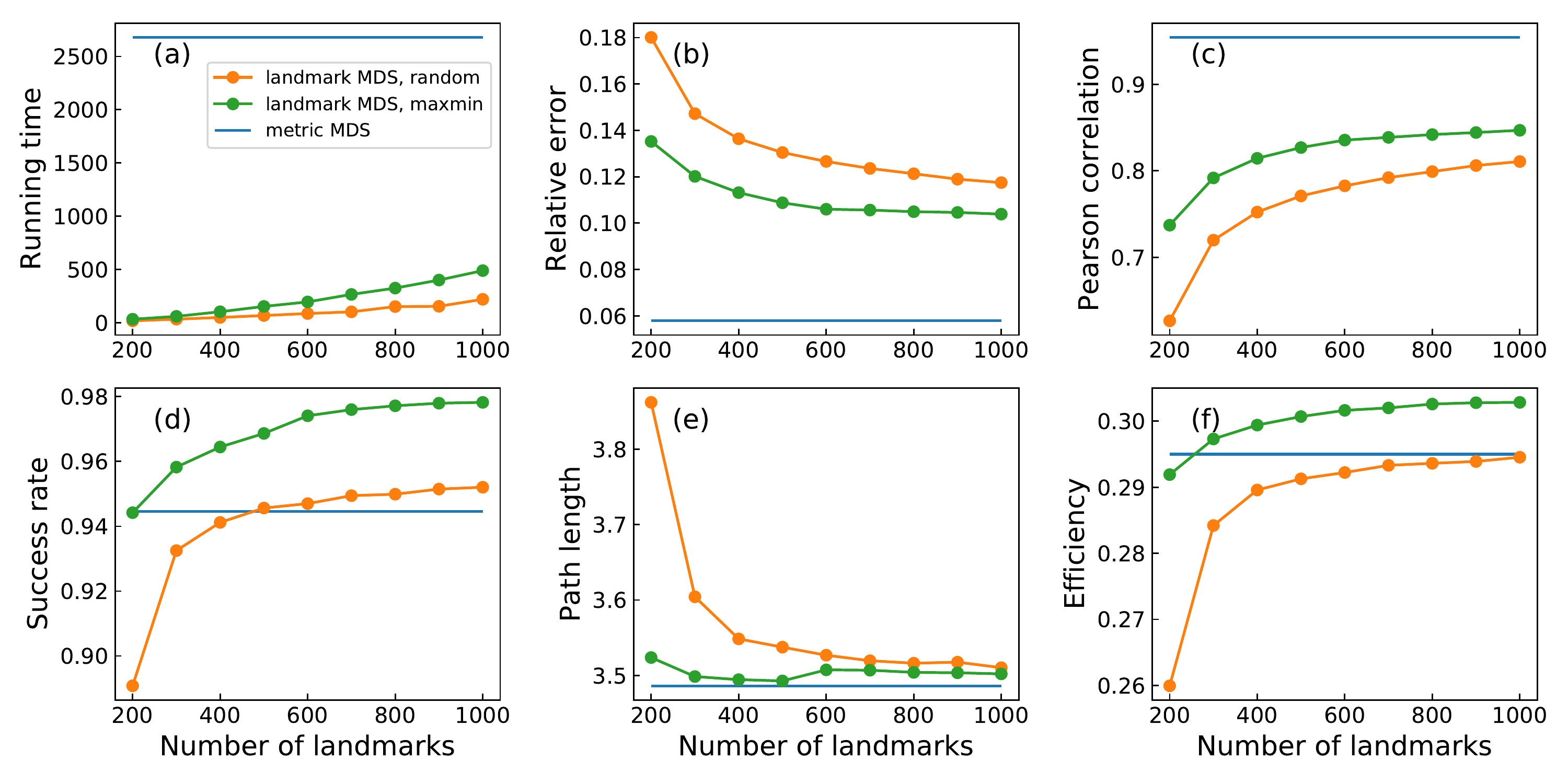}
\caption{\label{fig:sm_landmark_mds_performance_l} The performance of the landmark MDS algorithm on the IPv6 Internet network. (a) Running time, (b) relative error, (c) Pearson correlation, (d) greedy routing success rate, (e) greedy routing average path length, and (f) greedy routing efficiency as functions of the number of landmarks. The embedding dimension is $d = 100$.}
\end{figure*}

\begin{figure*}[!htb]
\includegraphics[width=0.85\textwidth]{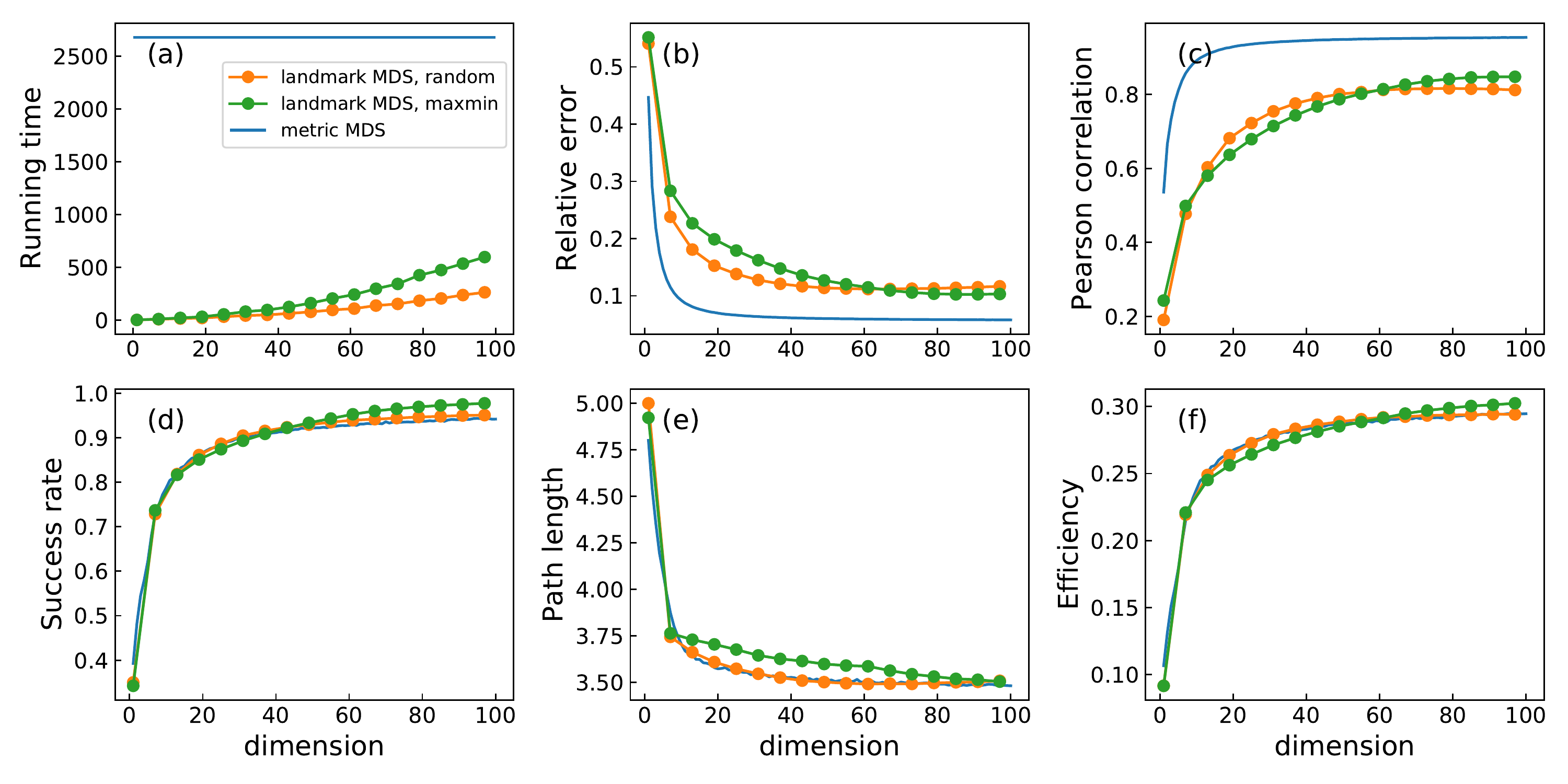}
\caption{\label{fig:sm_landmark_mds_performance_d} The performance of the landmark MDS algorithm on the IPv6 Internet network. (a) Running time, (b) relative error, (c) Pearson correlation, (d) greedy routing success rate, (e) greedy routing average path length, and (f) greedy routing efficiency as functions of the embedding dimension $d$. The number of landmarks is $l = 10d$.}
\end{figure*}



A different landmarks selection strategy in step 1 may affect the performance of landmark MDS. Here we compare two different landmark selection methods, random selection and Maxmin. Maxmin randomly picks the first landmark node $l_1$. Then, for $i \in \{2, ...,l \}$, the $i$-th picked landmark maximizes, over the remaining untouched nodes, the minimum shortest path length distance to any of the existing landmarks

\begin{equation}
    l_i = \underset{v \in V \backslash \{ l_1, ..., l_{i-1} \}}{\text{argmax}} \: \underset{l \in \{ l_1, ..., l_{i-1} \}}{\text{min}} \, D_{vl} ,
\end{equation}
where $V$ represents the set of all the nodes in the network.
The cost of using Maxmin instead of random selection amounts to $O(lN)$ extra operations.

In Fig.~\ref{fig:sm_landmark_mds_performance_l} and Fig.~\ref{fig:sm_landmark_mds_performance_d}, we compare the performance of landmark MDS with random selection and Maxmin methods on the IPv6 network ($N = 5143$). Metric MDS is used as the baseline.
For a systematic comparison, we report running time, relative error, Pearson correlation, greedy routing success rate, average path length, and efficiency in both figures. 

Figure.~\ref{fig:sm_landmark_mds_performance_l} focuses on the effect of landmark number. The range of landmark number is  $l \in [2d, 10d]$.
For both landmark selection approaches, $l = 4d$ is sufficient to get a respectable low-dimensional representation of the network. 
In Fig.~\ref{fig:sm_landmark_mds_performance_d}, we fix the number of landmarks with $l = 10d$ and study the effect of embedding dimension by varying $d \in [1, 100]$. 
Results show that while Maxmin takes more time than random selection, the performance is very close.
Therefore, we use the random selection approach with $l = 1000$ for landmark MDS in this paper.

\subsection{Computational and space complexity}

The naive version of the metric MDS algorithm has high computational and space complexity. The landmark MDS reduces the time and space complexity significantly when the number of landmarks is small, i.e., $l \ll N$.
Here we analyze the complexity of both algorithms.

To store the distance matrix, metric MDS requires $O(N^2)$ space while landmark MDS has a space complexity of $O(lN)$, where $N$ is the size of network and $l$ is the number of landmark nodes.
The computational complexity of MDS contains two parts:

\begin{itemize}
    \item[1.] Calculating distance matrix. Metric MDS requires $O(CN^2)$ as compared to $O(ClN)$ for landmark MDS, where $C$ is the cost to compute each entry of the distance matrix. For example, $C$ equals $O(L \text{log}N)$ with Dijkstra's algorithm, where $L$ is the number of edges in a network.
    \item[2.] Finding the embedding coordinates. The cost of metric MDS is dominated by the majorization of the stress function, i.e., the SMACOF algorithm. The time complexity of per iteration of SMACOF is at least $O(N^2)$. The dominating costs of landmark MDS are as follows: $O(l^2)$ for finding the embedding coordinates of landmarks and $O(dlN)$ for calculating the coordinates of nonlandmarks using the distance-based triangulation, where $d$ is the dimension of the embedding space.
\end{itemize}

In summary, metric MDS has computational complexity $O(CN^2 + N^2)$ and space complexity $O(N^2)$, while landmark MDS has computational complexity $O(ClN + dlN + l^2)$ and space complexity $O(lN)$.

We use landmark MDS for networks with more than 7,000 nodes in this work.


\section{Networks}

\subsection{Network data}
\label{appendix:networks}

We use both synthetic and real-world networks in this paper.
All networks are unweighted and undirected.
The nine real-world networks include two human brain networks, two transportation networks, one P2P network, and four snapshots of the Internet.
The synthetic networks include the following:

\begin{itemize}
    \item[(1)] \textit{Poisson networks}: They are generated by the configuration model with Poisson degree distribution. The size of network $N$ and average degree $\langle k \rangle$ are two tunable parameters.
    \item[(2)] \textit{$m$-ary trees}: These are rooted trees with each node having no more than $m$ children. We control both the parameter $m$ and the size of the tree $N$.
    \item[(3)] \textit{Popularity-similarity-optimization model (PSOM)} \cite{Papadopoulos2012Popularity, papadopoulos2015network_1}: PSOM is a growing network model that assumes nodes to be connected depending on their distance in a hidden hyperbolic space.  The parameters of PSOM are average degree $\langle k \rangle$, exponent $\gamma$ of the power-law degree distribution $P(k) \sim k^{-\gamma}$, and temperature $T$. Temperature $T$ controls the average clustering in the network, which is maximized at $T = 0$ and decreases to zero when $T$ approaches 1.
\end{itemize}

Details of the real-word networks and synthetic network instances can be found in Table~\ref{tab:network_information}.

\subsection{Network processing for visualization}
\label{appendix:networks_2d_map}

For clarity, only parts of the European Road (ER) network~\cite{Subelj2011} and the Air Transportation (AT) network~\cite{Guimera2005} are visualized in Fig.~\ref{fig:europe_2d_map}.
The original ER graph has $N=1,039$ nodes and $L=1,305$ edges.
The nodes represent European cities and an edge between two nodes represents a road connecting them.
We remove nodes corresponding to cities in Asian countries (Turkmenistan, Uzbekistan, Tajikistan, Kyrgyzstan, Iran, Syria, and Iraq) and transcontinental countries (Turkey, Georgia, Armenia, Kazakhstan, and Azerbaijan) as well as two nodes located in Kosovo, because the country is not represented in the AT network. 
In the end, we obtain a subgraph of ER with $N=859$ nodes and $L=1,098$ edges.

The original AT network has $N=3,618$ nodes and $L=14,142$ edges.
The nodes represent cities around the world.
A connection between two nodes indicates the existence of at least one flight between them from November 1st to 7th, 2000.
We obtain the European Air Transportation (EAT) network with $N=506$ nodes and $L=2,382$ edges by keeping the nodes located in countries that appear in the ER network.

The sub-graphs are only used in visualizing Fig.~\ref{fig:europe_2d_map}.
The original networks are used elsewhere.

\section{Relationship between the geometric distance of nodes in MDS embedding space and network closeness centrality}
\label{appendix:relation_to_centrality}

\begin{figure*}[!htb]
\begin{center}
\includegraphics[width=0.98\textwidth]{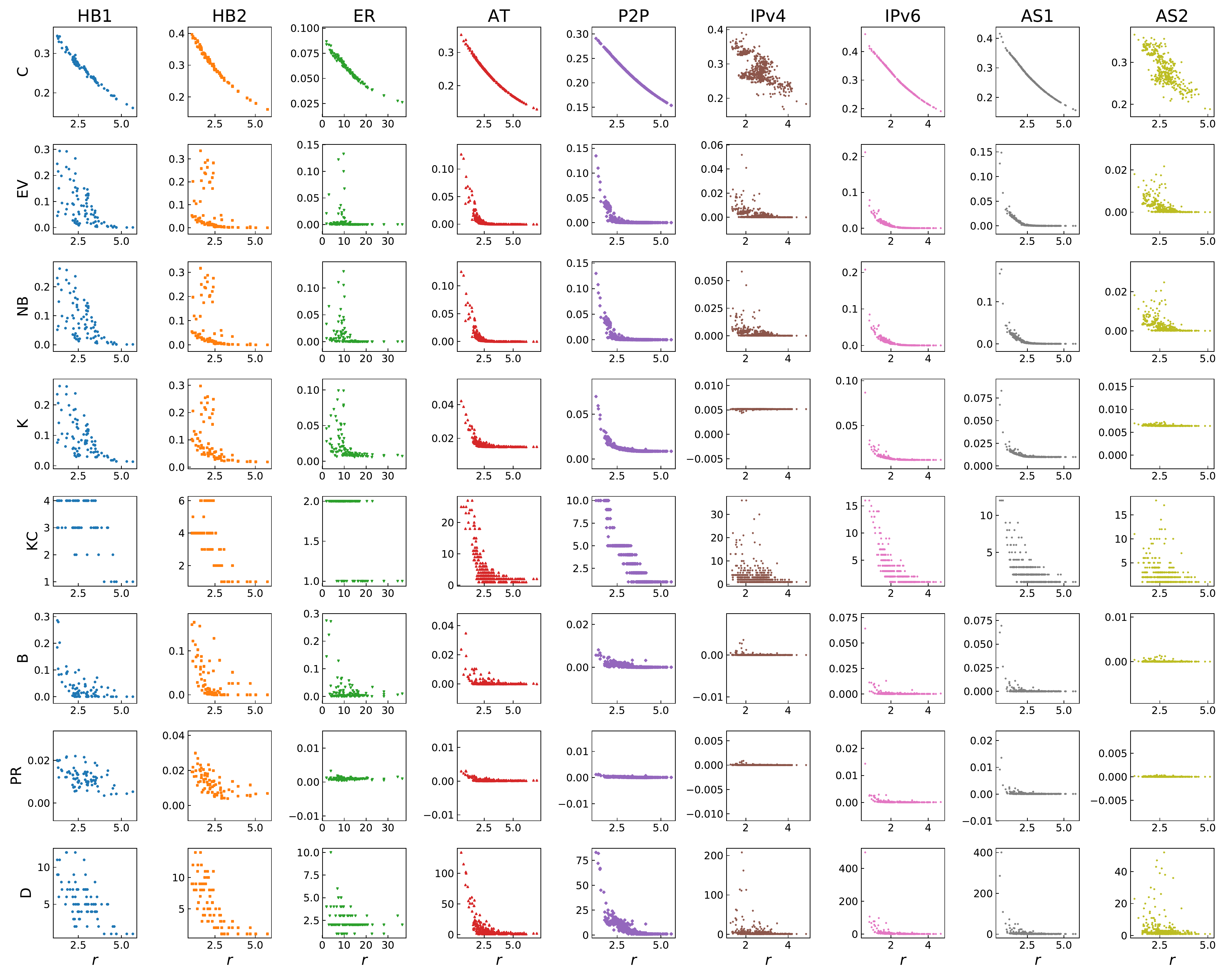}
\end{center}
\caption{\label{fig:sm_compare_centralities_relation}Pairwise relation between different centralities and the distance of a node from the geometric center in MDS embedding space ($r$) for different real-world networks. Network centralities considered  are closeness (C), eigenvector (EV), nonbacktracking (NB), Katz (K), k-core (KC), betweenness (B), PageRank (PR), and degree (D).
The embedding dimension is 100. The results for IPv4 and AS2 are obtained by landmark MDS with the number of landmarks $l = 1,000$. 
For clarity, we show only 10\% of all nodes if the network size is larger than 1,000.}
\end{figure*}

The network closeness centrality of node $i$ equals the inverse of the average shortest path distance of the node from all nodes in the network,
\begin{equation}
C_i = \frac{N}{\sum_j \, D_{ij}} \; .
    \label{eq:closeness}
\end{equation}
A perfect MDS embedding means that
\[
|| \mathbf{x}_i - \mathbf{x}_j || = a \, D_{ij}
\]
for all pairs of nodes $i, j$, with $a >0$ an arbitrary constant. We can write that 
\[
\begin{array}{ll}
N \, C_i^{-1} & = \sum_j \, D_{ij} = a^{-1} \sum_j || \mathbf{x}_i - \mathbf{x}_j || 
\\
& \leq a^{-1} \sum_j || \mathbf{x}_i || + a^{-1} \sum_j || \mathbf{x}_j || 
\end{array}\; ,
\]
where we used the triangle inequality. The center of the embedding corresponds to the origin of the reference frame. We have that $\sum_j || \mathbf{x}_i || = N || \mathbf{x}_i || = N \, r_i$, with $r_i$ distance of node $i$ from the geometric center of the embedding. Also, we can write that $\sum_j || \mathbf{x}_j || = N \, \langle r \rangle$ with $\langle d \rangle$ the average value of the distance of points from the geometric center. We can finally write the inequality
\begin{equation}
    C_i \geq \frac{a}{ r_i + \langle r \rangle } \; .
    \label{eq:closeness2}
\end{equation}

Our numerical results of Fig.~\ref{fig:compare_centralities} and of Fig.~\ref{fig:sm_compare_centralities_relation}
indicate that the inequality of Eq.~(\ref{eq:closeness2}) is tight in all networks considered in this paper. Other node centrality metrics also correlate with the distance of a node from the geometric center of the embedding. However, results seem network dependent.

\section{Evaluation metrics for MDS embedding}

\subsection{Relative error}
\label{appendix:relative_error}

We quantify the ability of MDS to preserve network proximity by estimating the relative error

\begin{equation}
E = \min_{\gamma >0} \left[ \frac{2}{N(N-1)} \sum_{i>j} \, \frac{ |D_{ij} - \gamma X_{ij}| } { D_{ij}} \right] \; ,
\label{eq:rel_error}
\end{equation}
where, for brevity of notation, we used $X_{ij} = ||\textbf{x}_i - \textbf{x}_j||$. Low values of $E$ correspond to good embeddings. Specifically, $E=0$ is obtained for an embedding that perfectly preserves the pairwise distance for all node pairs.

Note that for large networks it's infeasible to test every possible pair of nodes. For a network with $N > 500$, we choose $10^5$ random pairs of nodes to approximate the relative error. 

\subsection{Pearson correlation coefficient}
\label{appendix:pearson_correlation}

\begin{figure}[!htb]
\includegraphics[width=0.48\textwidth]{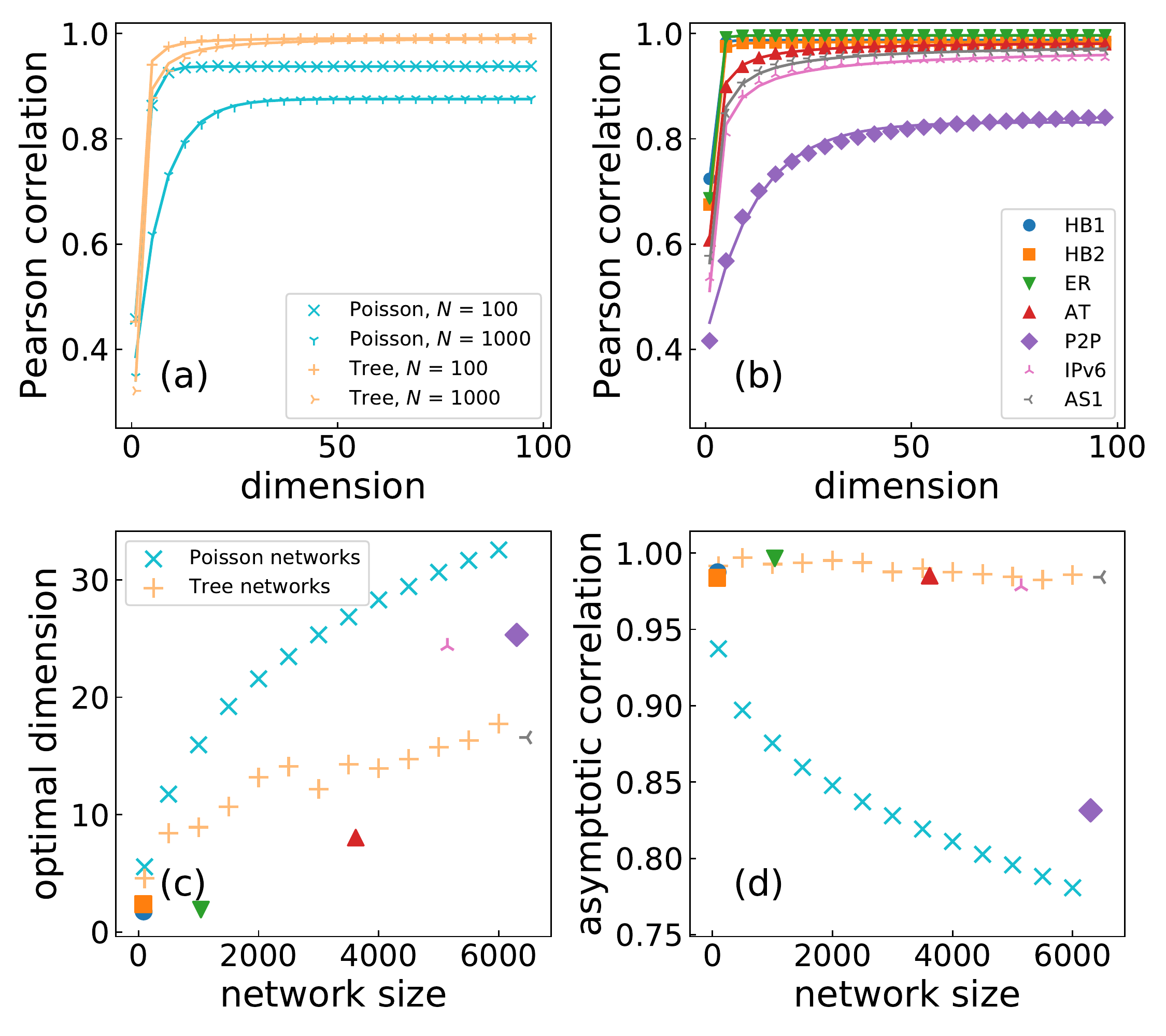}
\caption{\label{fig:sm_performance_correlation_MDS} Same as Fig.~\ref{fig:performance_mds}, but with the Pearson correlation $\rho$ between the shortest path length and distance in the embedding space as the evaluation metric.
For Poisson networks and P2P network, the continuous curves are obtained by fitting data points with exponential function $\rho(d) = \rho_{\infty} - s e^{\beta d}$. For tree networks and other real-world networks, the data points are fitted with power-law function $\rho(d) = \rho_{\infty} - s d^{\alpha}$. The optimal dimension is also calculated at the accuracy level $\epsilon = 0.05$.}
\end{figure}

The Pearson correlation coefficient between the pairwise shortest path length in the original networks and distance in the embedding space can be an alternative metric to measure how well the MDS embedding preserves the distance. The relative error and Pearson correlation values are all in the interval $[0, 1]$. For perfect embedding, the relative error should be zero and the Pearson correlation should be one. 

Figure.~\ref{fig:sm_performance_correlation_MDS} replicates Fig.~\ref{fig:performance_mds}, but with the Pearson correlation coefficient as the evaluation metric.
Similar to relative error, the results here also quickly saturate to an asymptotic value as the embedding dimension increases. The optimal dimensions obtained using Pearson correlation coefficients are slightly different from the ones calculated using relative error, but the values are still very small for all networks considered.

\section{Estimating the optimal embedding dimension}
\label{appendix:optimal_dimension}

To estimate the optimal dimension of MDS embedding for different networks, we use the method introduced in Ref.~\cite{gu2020defining}.
Assuming the plateau value $E_{\infty}$ of the relative error $E$ corresponds to the best geometric description that the embedding algorithm can achieve, the optimal dimension $d_o(\epsilon)$ at accuracy level $\epsilon$ is defined as

\begin{equation}
    d_o(\epsilon) = \text{arg} \, \underset{d}{\text{min}} (E(d) - E_{\infty} < \epsilon) \; , \label{eq:optimal_dimension}
\end{equation}
i.e., the minimal $d$ value such that the difference between $E(d)$ and the optimum $E_{\infty}$ is at most $\epsilon$.
Our numerical tests indicate that $E(d)$ can be well described by the function

\begin{equation}
    E(d) = E_{\infty} + s d^{- \alpha}  \;.
    \label{eq:parametric}
\end{equation}
We fit data points to Eq.~(\ref{eq:parametric}) and obtain the best estimates of the parameters $\hat{E}_{\infty}, \hat{s}$ and $\hat{\alpha}$. 
The best estimate of the optimal embedding dimension $\hat{d}_o(\epsilon)$ is calculated as

\begin{equation}
    \hat{d}_o(\epsilon) = \left( \frac{\hat{s}}{\epsilon} \right)^{1/\hat{\alpha}} \; .
    \label{eq:opt_estimate}
\end{equation}
The estimated optimal dimensions and the asymptotic relative error $\hat{E}_{\infty}$ are shown in Table~\ref{tab:network_information}.

\section{Greedy routing}
\label{appendix:greedy_routing}

\begin{figure}[!htb]
\includegraphics[width=0.4\textwidth]{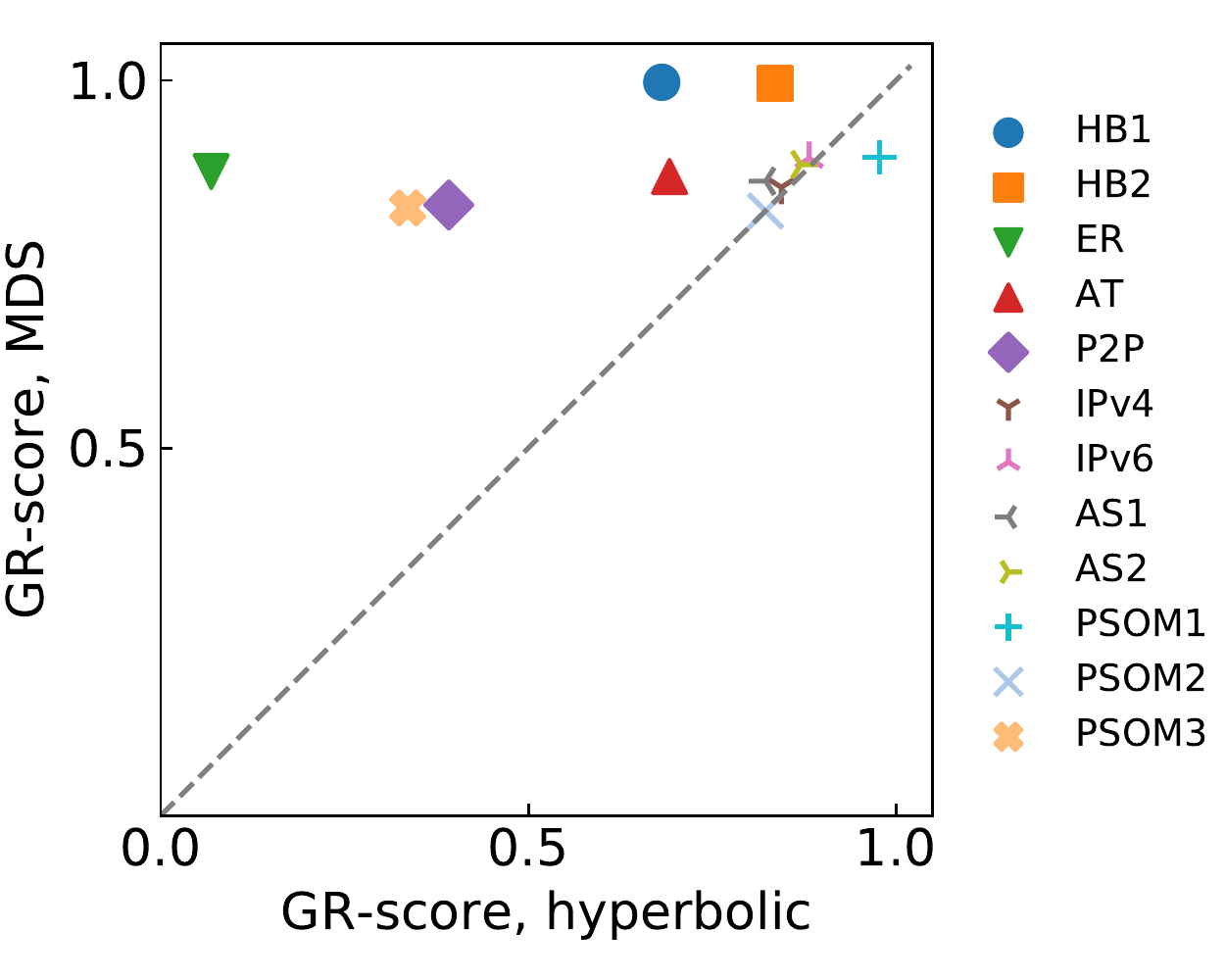}
\caption{\label{fig:sm_comparison_gr_score} Same analysis as in Fig.~\ref{fig:comparison_greedy_routing_rate}, but the performance is measured by the GR-score.}
\end{figure}

We randomly select $10,000$ source-target pairs. Starting from the source node, the packet tries to reach the target node using the greedy strategy described in the main text. Two outcomes are possible: (1) the packet reaches its destination in $R$ steps or (2) the packet fails to reach its target. To evaluate the performance of greedy routing, three metrics are used:
\begin{itemize}

\item[1.] The success rate $p_s$ defined as the ratio of correctly delivered packets over total number of packets considered~\cite{boguna2009navigability}.

\item[2.] Efficiency $\eta = p_s \langle 1/R \rangle$, where $\langle 1/R \rangle$ represents the mean value of the inverse of the path length $R$ obtained for each packet successfully delivered~\cite{faqeeh2018characterizing}.

\item[3.] GR-score $ = \frac{2}{N(N-1)} \sum_{i > j} \frac{D_{ij}}{R_{ij}}$, where $D_{ij}$ is the shortest path length from $i$ to $j$ in graph and $R_{ij}$ is the greedy routing path length from $i$ to $j$. The GR-score considers all successful and unsuccessful delivering. When greedy routing is unsuccessful, $R_{ij}$ is infinite and $D_{ij}/R_{ij} = 0$~\cite{muscoloni2017machine}.

\end{itemize}

In the main text, we use the success rate and efficiency to evaluate the performance of greedy routing. Here we implement the GR-score for the evaluation. As shown in Fig.~\ref{fig:sm_comparison_gr_score}, the result of the GR-score is similar to the success rate in main text. 

\begin{figure*}[t]
\begin{center}
\includegraphics[width=0.98\textwidth]{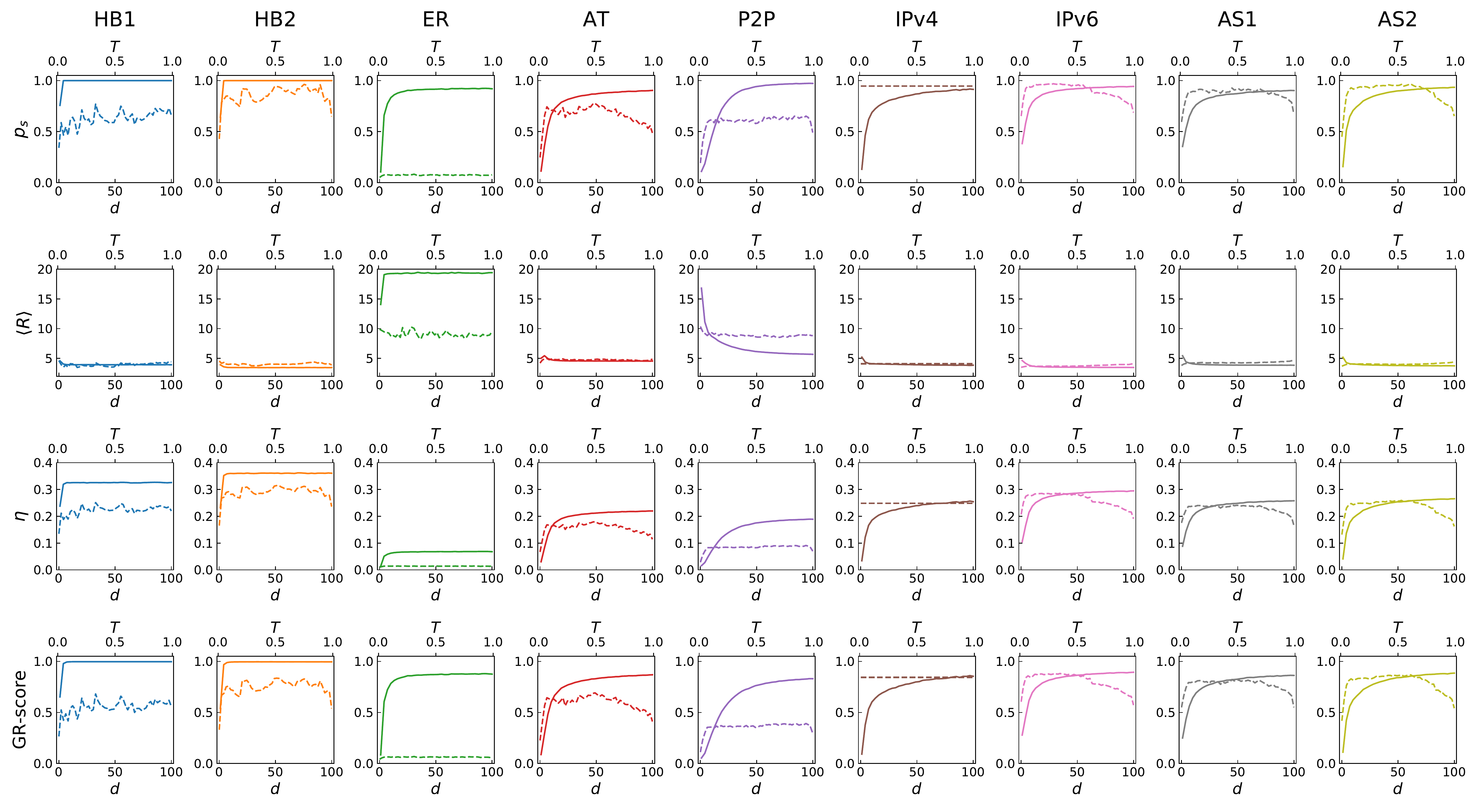}
\end{center}
\caption{\label{fig:sm_greedyrouting} The comparison of greedy routing success rate $p_s$, average path length $\langle R \rangle$, efficiency $\eta$, and GR-score on MDS and hyperbolic embedding for different real-world networks. The solid lines represent results based on MDS embedding, and the dashed lines are the results based on hyperbolic embedding.}
\end{figure*}

In our tests, we compare the performance of MDS and hyperbolic embedding in guiding greedy routing. Both methods have a tunable parameter that can affect the navigation performance. In MDS embedding, the parameter is the dimension $d$ of the embedding space. We consider discrete values in $[1, 100]$ for $d$. In hyperbolic embedding, the parameter is temperature $T$. We try different values in $[0, 1]$ for $T$. Results in Fig.~\ref{fig:comparison_greedy_routing_rate} are obtained using the parameter values that maximize the metrics for both methods.
Detailed results are shown in Fig.~\ref{fig:sm_greedyrouting}.

\section{Susceptible-Infected processes on networks}

\subsection{Susceptible-Infected model}
\label{appendix:si_model}

In the susceptible-infected (SI) model, the state of the individuals in the network is either susceptible or infected. Susceptible individuals do not carry the disease, but they can be infected. Infected individuals carry the disease, and they can spread it to susceptible individuals. The rate of infection is $\beta$.

We use the  Gillespie algorithm to simulate SI dynamics~\cite{gillespie1977exact}. The steps of the algorithms are  as follows:

\begin{itemize}
    \item [1.] At time $t = 0$, randomly select one infected individual, and all other individuals are susceptible.
    
    \item [2.] Determine the time interval $\Delta t$:
    \begin{equation}
        \Delta t = \frac{- \log (u)}{\beta \overset{\frown}{SI}},
    \end{equation}
    where $u$ is a random number extracted from the uniform distribution with support in the interval $(0, 1)$, and $\overset{\frown}{SI}$ is the number of susceptible-infection pairs.
    
    \item [3.] Randomly select a susceptible-infection pair and let the susceptible individual be infected. Increase  time from $t$ to $t + \Delta t$.
    
    \item [4.] Iterate step 2 and step 3 until all susceptible individuals become infected.
\end{itemize}

For each network, we simulate the SI process $N$ times to get the average arrival time of infection for every node where $N$ is the network size. 

\subsection{Results on additional networks}
\label{appendix:si_process}

\begin{figure*}[!htb]
\begin{center}
\includegraphics[width=0.98\textwidth]{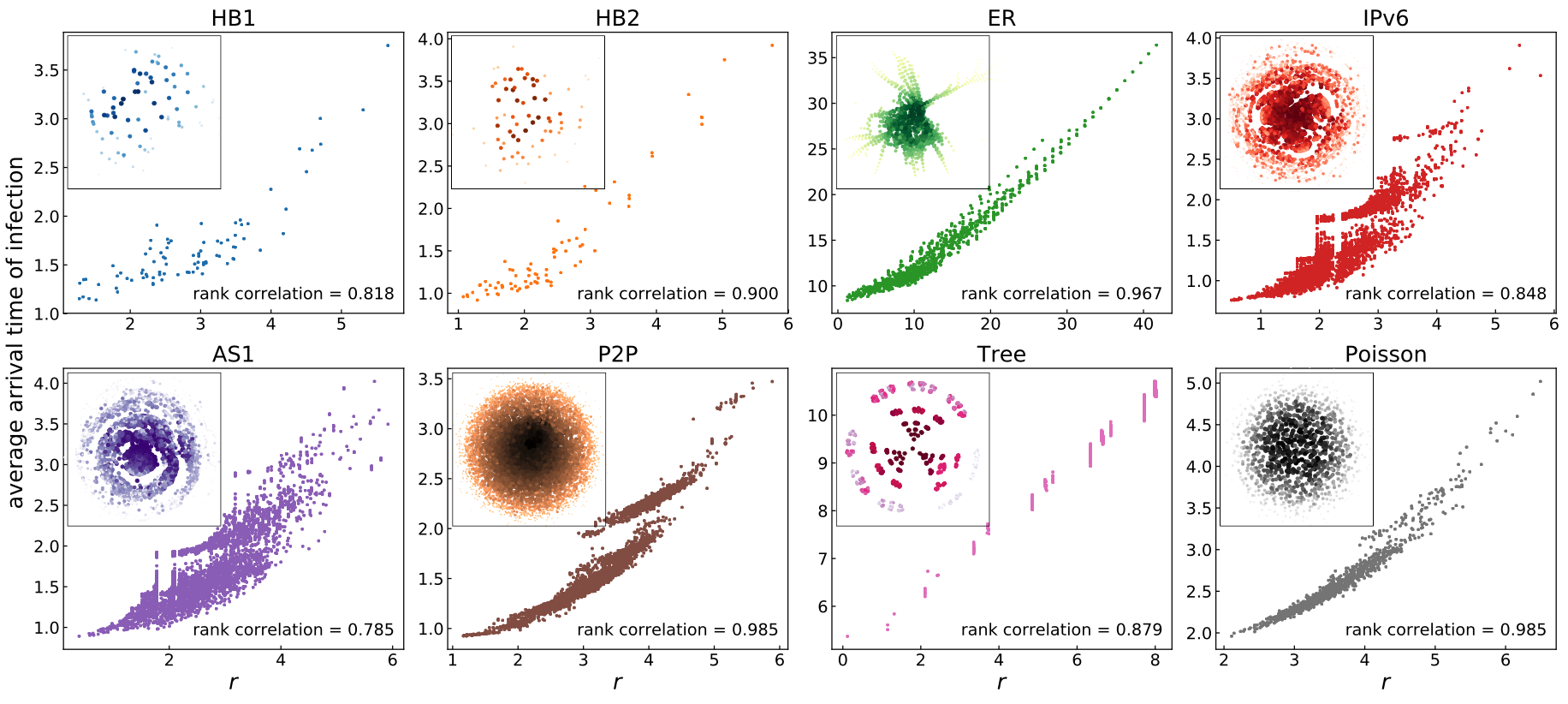}
\end{center}
\caption{\label{fig:sm_time_distance_all}
Scatter plot of the average infection arrival time against the distance of nodes from the geometric center of the MDS map for different real-world networks and synthetic networks. Each point in the plot represents a node in the network. The average infection arrival time is calculated based on $N$ independent simulations of the susceptible-infected model with a random seed each time, where $N$ is the network size. The network is embedded with $d=100$. For illustrative purposes only, we display in the inset how the spreading looks like in a $2$-dimensional MDS map. Nodes with larger size and darker color are infected earlier.}
\end{figure*}

In the main text, we show the relation between average infection arrival time and the distance of nodes from the geometric center of the MDS map, and the spreading pattern for the air transportation (AT) network (Fig.~\ref{fig:time_distance_relation_AT}). Here, we report the results for other real-world and synthetic networks (Fig.~\ref{fig:sm_time_distance_all}). For all networks tested, the rank correlations are very high, and clear concentric wave patterns in the two-dimensional representations of the networks can be observed in the insets of Fig.~\ref{fig:sm_time_distance_all}.

\bibliography{main}

\end{document}